# НЕКОТОРЫЕ СВОЙСТВА $\lambda$-СОЛИТОНОВ

В.В. МАХРО*), И.Г. МАХРО**)
*)Иркутский государственный университет, филиал в г. Братске
665729 Братск, ул. Ленина, 34
**) Братский государственный университет
665709 Братск, ул. Макаренко, 40

## Some properties of $\lambda$ - solitons

We investigate some dynamical parameters of $\lambda$ - solitons which arises in the family of implicit difference schemes for the diffusion equation with fractional difference operator $\frac{\partial^2}{\partial x^2}$ and imaginary diffusion coefficient. We suppose that such schemes may correspond to fractional diffusion equation with imaginary diffusion coefficient.

Ранее мы сообщали [1, 2] об обнаружении солитоноподобных решений в семействе конечно-разностных схем, строящихся на основе неявной схемы Эйлера вариацией параметра $\lambda$ в пространственном дифференциально-разностном операторе

$$\frac{\partial^2 u}{\partial x^2} = \frac{1}{h^2}\left(u_{j+1,m+1} - \lambda u_{j,m+1} + u_{j-1,m+1}\right)$$

где $h$ - пространственный шаг сетки, для уравнении диффузии

$$\frac{\partial u}{\partial t} = C\frac{\partial^2 u}{\partial x^2}$$

с чисто мнимым коэффициентом $C$. В частности, было обнаружено, что устойчивые солитоноподобные структуры формируются для всех $\lambda \in (0, 2)$.

В настоящей заметке мы приводим новые результаты исследования таких структур (здесь, и далее – $\lambda$-солитонов). Основные динамические параметры $\lambda$-солитонов исследовались при нулевых граничных условиях, стабильных начальных условиях $u(x,0) = 1$, и при вариации параметров $C$ и $\lambda$.

Нами установлено, что возникновение $\lambda$-солитонов носит пороговый характер, то есть они возникают при наличии градиента в начальных условиях, составляющего по порядку 0.001 (в зависимости от величины $\lambda$ это число слабо меняется).

Численными методами мы подробно исследовали скорость движения $\lambda$-солитонов, как функцию параметра $\lambda$. Было установлено, что зависимость скорости солитонов от $\lambda$ описывается монотонными функциями, представленными на *рис. 1*.

Нами также было изучено изменение высоты солитонов во время их движения как функции $C$ в широком интервале значений (от 2.8 до 0.5). Результаты представлены на *рис. 2*. Пики на графике соответствуют столкновению солитонов, распространяющихся навстречу друг другу из противоположных краев области интегрирования, а также их отражению от границ интервала интегрирования (при моделировании этот интервал брался равным 1).

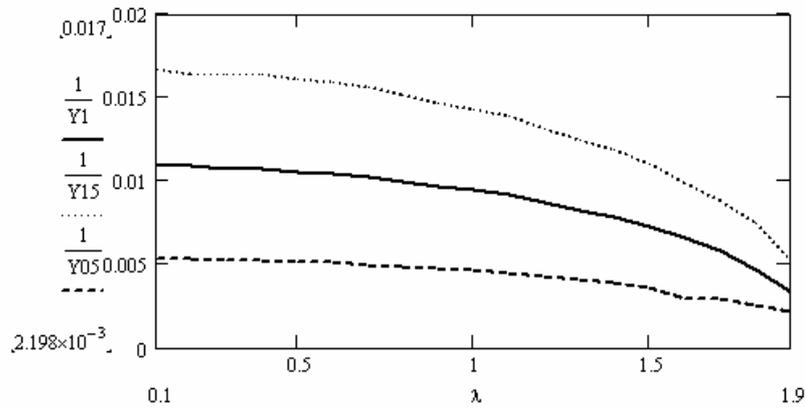

*Рис. 1*. Зависимость относительных скоростей $\lambda$-солитонов для нескольких значений параметра $C$: 1/Y1 – 1*i*, 1/Y15 – 1.5*i*, 1/Y05 – 0.5*i*.

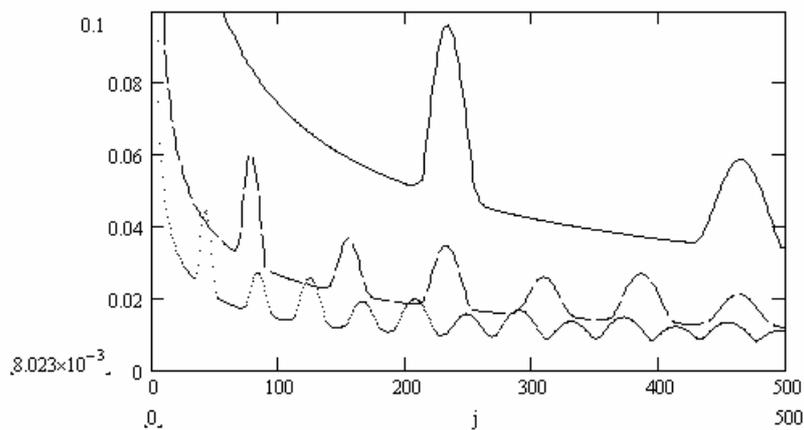

*Рис. 2.* Изменение высоты $\lambda$ - солитонов в процессе движения, $j$ – временной параметр, $C$ (сверху вниз 2.8$i$, 1.5$i$, 0.5$i$).

По-прежнему требует дальнейшего исследования вопрос об интерпретации полученных решений и их соответствия реальным дифференциальным уравнениям. В качестве рабочей гипотезы мы обсуждаем предположение о том, что исследуемое семейство разностных схем соответствует уравнению диффузии (в мнимом времени, то есть уравнению Шредингера в реальном) с дробными производными в смысле Грюнвальда - Летникова:

$$\frac{u_{j,m+1} - u_{j,m}}{\Delta t} = K_{\gamma\ -0} D_t^{1-\gamma} \left( \frac{u_{j-1,m+1} - 2u_{j,m+1} + u_{j+1,m+1}}{(\Delta x)^2} \right).$$

Фактически, доказательство данного утверждения будет означать получение мощного инструмента для решения широкого класса задач математической физики.

Однако и сейчас полученные результаты могут представлять значительный интерес для различных фундаментальных и прикладных задач.